\documentclass[aps,prb,amsmath,twocolumn,amssymb,titlepage,superscriptaddress,showpacs]{revtex4-1}
\usepackage[english]{babel}
\usepackage{graphicx}
\usepackage{transparent}
\usepackage{amsmath}
\usepackage{amssymb}
\usepackage{epstopdf}
\usepackage{amsfonts}
\usepackage{bm}
\usepackage{mathdots}
\usepackage{times}
\usepackage{pdfpages}
\usepackage{mathtools}
\usepackage{stmaryrd}
\usepackage{tabularx}
\usepackage{hhline}

\makeatletter
\AtBeginDocument{\let\LS@rot\@undefined}
\makeatother

\newcommand{\ave}[1]{\langle #1 \rangle}
\newcommand{\bolds}[1]{\boldsymbol #1}

\begin{document}
\title{Momentum-dependent relaxation dynamics of the doped repulsive Hubbard model}
\author{Sharareh Sayyad}
\affiliation{Institute for Solid State Physics, University of Tokyo, Kashiwanoha,
Kashiwa, Chiba 277-8581, Japan}
\author{Naoto Tsuji}
\affiliation{RIKEN Center for Emergent Matter Science (CEMS), Wako 351-0198, Japan}
\author{Abolhassan Vaezi}
\affiliation{Department of Physics, Stanford University, Stanford, CA 94305, USA}
\affiliation{5Stanford Center for Topological Quantum Physics, Stanford University, Stanford, CA, USA}
\author{Massimo Capone}
\affiliation{International School for Advanced Studies (SISSA), Via Bonomea 265, I-34136 Trieste, Italy}
\affiliation{CNR-IOM Democritos, Via Bonomea 265, I-34136 Trieste, Italy}
\author{Martin Eckstein}
\affiliation{Department of Physics, University of Erlangen-N\"{u}rnberg, 91058 Erlangen, Germany}
\author{Hideo Aoki}
\affiliation{National Institute of Advanced Industrial Science and Technology (AIST), Tsukuba 305-8568, Japan}
\affiliation{Department of Physics, University of Tokyo, Tokyo 113-0033, Japan}

\pacs{71.10.Hf, 71.30.+h, 71.27.+a, 74.72.Gh, 74.72.Ek}

\begin{abstract}

We study the dynamical behavior of doped electronic systems subject to a global ramp of the repulsive Hubbard interaction. We start with formulating a real-time generalization of the fluctuation-exchange approximation.  
Implementing this numerically, we investigate the weak-coupling regime
of the Hubbard model both in the electron-doped and hole-doped regimes. The results show that both local and nonlocal (momentum-dependent) observables evolve toward a thermal state, although the temperature of the final state depends on the ramp duration and the chemical doping. 
We further reveal a momentum-dependent relaxation rate of the distribution function in doped systems, and trace 
back its physical origin to the anisotropic self-energies in the momentum space.

\end{abstract}
\maketitle

\section{Introduction}

There are increasing fascinations toward 
optimizing and controlling the properties of correlated 
systems in both experimental and theoretical disciplines. 
In particular, driving a system out of equilibrium 
by applying time-dependent modulations is a 
powerful avenue for realizing new quantum states.  
When the time-translational symmetry is broken in a nonequilibrium pathway, the initial thermal state 
can transform into a different state in a nonthermal fashion, since the time-dependent drive incorporates the excitations of the correlated system. At short times, different non-thermal transients can emerge depending on the initial correlations and the preparation protocol. In the long-time limit, it is intriguing to 
see whether the system retrieves a thermal state or whether new correlations are built up which realize a long-lived nonthermal state. Achieving comprehensive insights requires, as discussed in the literature, incorporating details of the system including the interplay of various degrees of freedom\cite{Giannetti2016,Perfetti2006, Konstantinova2018, Smallwood2014, Smallwood2012} lattice structures,~\cite{Demsar2003, Johnson2017} and doping concentrations~\cite{Torchinsky2011}. 

In low dimensional systems where strong nonlocal quantum correlation play a significant role, such as, e.g. layered correlated oxides or heterostructures, the doping level is a key control parameter, which leads to different regimes where different scattering mechanisms dominate.~\cite{Chan2015, Stewart2017} This should leave fingerprints on the measured observables. In particular, the time-resolved optical conductivity after an excitation often shows a single or double exponential decay,~\cite{Hu2014} measuring Fermi-surface properties, i.e., the momentum-dependent distribution function, and reveal more detailed information and display distinct temporal relaxation responses at different points on the Fermi surface.~\cite{Smallwood2012,Smallwood2014,Graf2011,Ishida2016,Cortes2011} Understanding the details of these is a formidable challenge due to the interplay between various degrees of freedom in a small energy range~\cite{Keimer2015, Stewart2017}, and dealing with the complexity of, usually, 2D layered structures of these multi-band~ (-orbital) correlated materials. Nevertheless, it is 
imperative to understand the relaxation dynamics of doped systems, where the electron-electron interaction governs the physics. 

From a theoretical perspective, even when we focus on the on-site (Hubbard) repulsive 
interaction, investigation of the problem is quite 
challenging due to the nonequilibrium nature of the problem.

Incorporating spatially nonlocal correlations in two dimensions is computationally demanding as in nonequilibrium dynamical cluster approximation~(DCA)~\cite{Tsuji2014}, diagrammatic approaches~\cite{Bauer2015,Dasari2018}, and variational Monte Carlo method~\cite{Ido2017}, while the well-developed 
one-dimensional~\cite{Kollath2007, Bischoff2017} and infinite-dimensional~\cite{Aoki2014} nonequilibrium algorithms cannot directly treat two-dimensional systems. 
A large number of investigations have been devoted to understanding the long-time dynamics of fermionic systems after a global ramp of interaction parameter, at half filling.~\cite{Tsuji2014, Bauer2015, Eckstein2009, Tuji2013, Schiro2010} In these attempts, the long-time thermalization occurs for nearly all the weak electron-electron coupling regimes, which sometimes supersedes intermediate pre-thermalization plateaus.~\cite{Bauer2015, Eckstein2009, Schiro2010} 
Away from the half-filling, on the other hand, 
investigations of doped systems in infinite dimensions reveal that the observed sharp dynamical transition from an exponential relaxation in the weak-coupling regime to an oscillating behavior in the strong-coupling regime is smoothened into a crossover between these two regimes.~\cite{Schiro2010}    
In two dimensions, targeting half-filled systems reveal a momentum-dependent relaxation rate of the single-particle momentum distribution after a sudden quench of the electron-electron interaction.~\cite{Tsuji2014, Bauer2015}  
Thus the question remains as to 
(i) how these dynamical behaviors vary in doped systems,
(ii) what would be the effect of finite ramp durations 
(as opposed to sudden quenches), and
(iii) to what extent the relaxation dynamics depends 
on the momenta in the Brillouin zone.  

In the present work we precisely address this question in both the electron-doped and hole-doped systems. We employ an interaction ramp protocol on a two-dimensional system with repulsive electron-electron interaction. We formulate the real-time generalization of the fluctuation exchange approximation (FLEX)\cite{Bickers1991, Bickers1989, Kuroki1999, Oka2010}. It is known that this algorithm requires 
vertex corrections in the intermediate and strong-coupling regimes where the Mott insulator starts to emerge.~\cite{Ayral2015} Also, 
underdoped systems exhibit, at low temperatures, pseudogap physics which 
requires an extension of the FLEX.~\cite{Yanase2001}  We thus limit ourselves to the weak-coupling regime and at temperatures where pseudogap does not emerge.  We show that there exist a rapid local thermalization of the 
system, where the ramp duration determines the temperature. More importantly, using the nonequilibrium FLEX, we can analyze the {\it momentum-dependent evolution} of both single- and two-particle observables. Namely, our results exhibit a doping-dependent nodal-antinodal dichotomy in the relaxation rate of the single-particle momentum distributions.

This paper is organized as follows. We define the model Hamiltonian and present the proposed numerical algorithm in Sec.~\ref{sec:Ham}. In Sec.~\ref{sec:results} we present numerical results and discuss the underlying physics. Section~\ref{sec:conclusion} is devoted to a summary and conclusions of this work.

\section{Model and Method}\label{sec:Ham}

\subsection{Model Hamiltonian} 
The repulsive one-band Hubbard model on the square lattice is defined as
\begin{align}
 H =
&-
\sum\limits_{ij, \sigma} \big(
v_{|i-j|} 
c^{\dagger}_{i\sigma}
c_{j\sigma}
+ {\rm h.c.}
\big) \nonumber \\
&
+
\sum\limits_{i} U(t)
\left(n_{i\uparrow} -\frac{1}{2} \right)
\left(n_{i\downarrow} -\frac{1}{2} \right)
-\mu \sum\limits_{i,\sigma} n_{i\sigma} 
,\label{eq:Hubbard}
\end{align}
where $c^{\dagger}_{i \sigma}(c_{j \sigma})$ creates~(annihilates) an electron with spin $\sigma$ at site $i$. The band filling is set by the chemical potential $\mu$. Density of electrons, $n_{i \sigma}$, with opposite spins experience a local repulsive interaction~$U$. An electron hops from site $i$ to a neighboring site $j$ with the hopping amplitude of $v_{|i-j|}$, which is taken here up to the third-neighbors. 
We then have a band dispersion on the square lattice as
 \begin{align}
  \varepsilon_{\bolds{k}} =& -2 v_{1} \big[\cos(k_{x}) + \cos(k_{y})\big]
  -4 v_{2} \cos(k_{x})  \cos(k_{y}) \nonumber \\
  &-2 v_{3} \big[\cos(2k_{x}) + \cos(2k_{y})\big],\label{eq:disp}
 \end{align}
where $\bolds{k}=(k_{x},k_{y})$ is the two-dimensional momentum, and hopping parameters are here set to $v_{2}=-0.2v_{1}$, and $v_{3}=0.16 v_{1}$, relevant for the the copper-oxygen planes of high-temperature superconductors~\cite{Kitatani2014,Sakakibara2010}, see solid green line in Fig.~\ref{Fig:Gk}~(lower panel). We report our results in the energy~(time) unit of $v_{1}~(1/v_{1})$.

\begin{figure}[t]
\includegraphics[width=0.47\textwidth]{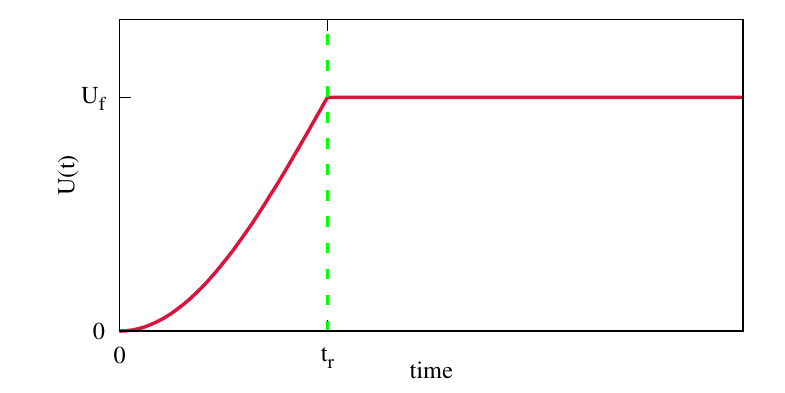}\\[-3mm]
\caption{
 Ramp up of the Hubbard interaction given in Eq.~\eqref{eq:Ut}. 
}
\label{Fig:UT}
\end{figure}

To explore the thermalization of doped quantum systems,  
we take the Hubbard model where the Hubbard interaction 
is switched on with a ramp (see Fig.\ref{Fig:UT}), 
\begin{equation}\label{eq:Ut}
 U(t)= 
 \begin{cases}
  U_{\rm f} \left[ 1-\cos(\frac{\pi t}{2 t_{\rm r}}) \right] \quad \text{for} \quad  t\leq t_{\rm r},\\
  U_{\rm f} \qquad \qquad \qquad \quad \, \text{for} \quad t>t_{\rm r},
 \end{cases}
\end{equation}
with $t_{\rm r}$  the ramp duration, and $U_{\rm f}$ the final Hubbard interaction. This protocol can directly be realized in cold-atom setups~\cite{Jo2009}, or can indirectly 
be realized  in solids  excited by a short few-cycle laser pulse~\cite{Ligges2018} in terms of the effective electron-electron interaction.   
As the FLEX is reliable in the weak-coupling regime, we set the final Hubbard interaction to $U_{\rm f} \leq 3$ for which $U_{\rm f}$ is much smaller than the band width ($=7.68$ for the 
present choice of the hopping parameters). The initial temperature $\beta=1/T=20$ is chosen such that the system 
is away from superconducting and antiferromagnetic phases.  

\subsection{Numerical method}

We perform the numerical investigations of our model in finite-temeprature paramagnetic model where no long-range order is present, and use, for nonequilibrium situations, 
the Schwinger-Keldysh~\cite{Keldysh1965, Kamenev2011} generalization of the FLEX~\cite{Bickers1991, Bickers1989, Kuroki1999}.  
Included Feynman graphs in this diagrammatic formalism are determined from a functional derivative of the Luttinger-Ward functional,~\cite{Baym1961} and thus the approach is a conserving approximation.~\cite{Bickers1991, Aryanpour2003} It has been well established that in the weak-coupling regime the results are qualitatively in good agreement with numerically exact quantum Monte-Carlo results.~\cite{Dahm1994} 
One technical detail is that the self-energy diagrams have been collected here with an assumption that the expectation value of the pair correlation is negligible and thus the anomalous contributions can be omitted. This particular choice implies that our formalism is applicable to studying normal phases as well as investigating the behavior of the system just at an instability such as superconductivity.

In the FLEX, the electron scattering incorporates the magnetic, density, and also singlet-pairing channels. 
Correspondingly, we consider the spin susceptibility ($\chi^s$), charge susceptibility ($\chi^c$), and, here, in particular, the particle-particle susceptibility ($\chi^{pp}$). The later is important since it enables us to 
examine superconducting instabilities through pair fluctuations represented 
by the particle-particle susceptibility, which is defined in terms of the pair operator, $\Delta_{\bolds{k}}(t)= c_{\bolds{k} \uparrow}(t) c_{-\bolds{k}\downarrow}(t)$, as
\begin{equation}
 \chi^{pp}_{\bolds{k}}(t,t') = -i \ave{ {\cal T}_{\cal C}  \Delta_{\bolds{k}}(t)  \Delta^{\dagger}_{-\bolds{k}}(t')},
\end{equation}
where $\cal T_{\cal C} $ is the time-ordering operator on the Keldysh contour $\cal C$.

The spin, charge, and particle-particle susceptibilities obey algebraic equations which are obtained by summing geometric series of ladder and bubble diagrams,
\begin{equation}
 \chi^{r}_{\bolds{q}}(t,t') = U(t) \chi_{\bolds{q}}^{0r}(t,t') U(t') +s_{r} \big[ \chi^{0r}_{\bolds{q}} * U* \chi^{r}_{\bolds{q}} \big] (t,t'), 
\end{equation}
where $r \in \{s,c,pp\}$ denotes the channel index, and $s_{r}$ has $s_{s}=s_{pp}=-1, s_{c}=1$. The crystal momentum $\bolds{q}$ resides on a $N_{k}\times N_{k}$ momentum grid associated with $\sum_{\bolds{q}} =N_{k}^{2}$. In nonequilibrium situations, every susceptibility 
has two time arguments due to the loss of time-translational invariance.  
 Time arguments $t,t'$ are defined on the Schwinger-Keldysh contour $\cal C$, and $*$ is the convolution integral on $\cal C$. $\chi^{0r}$ is the polarization function computed by
\begin{align}
 \chi_{\bolds{q}}^{0c/s}(t,t') &= \frac{\mathrm{i}}{N_{k}^{2}} \sum\limits_{\bolds{k}} G_{\bolds{k}+\bolds{q}}(t,t') G_{\bolds{k}}(t',t) ,\\
 \chi_{\bolds{q}}^{0pp}(t,t') &= \frac{\mathrm{i}}{N_{k}^{2}} \sum\limits_{\bolds{k}} G_{\bolds{k}+\bolds{q}}(t,t') G_{-\bolds{k}}(t,t'),
\end{align}
where $G$ is the interacting Green's function,
\begin{equation}\label{eq:Gint}
 G_{\bolds{k}}(t,t')= -i \langle {\cal T}_{\cal C}[c_{\bolds{k}}(t) c^{\dagger}_{\bolds{k}}(t') ]\rangle.
\end{equation}

This single-particle propagator satisfies the Dyson equation,
\begin{align}
 &\big( i \partial_{t} + \mu -\Sigma^{\rm H} - \varepsilon_{\bolds{k}} \big){\cal G}_{\bolds{k}}(t,t') =\delta_{\cal C}(t,t'),\nonumber\\
 &G_{\bolds{k}}(t,t') = {\cal G}_{\bolds{k}}(t,t') + \big[ {\cal G}_{\bolds{k}} * \Sigma_{\bolds{k}}^{\rm FLEX}  * G_{\bolds{k}} \big] (t,t'). \label{eq:Dysonint}
\end{align}
The local electron density is given as 
\[
n(t) = \frac{2}{N_{k}^{2}}  \sum_{\bolds{k}} {\rm{Im}} [G_{\bolds{k}}^{<}](t,t) , 
\]
where $G^{<}$ is the lesser component of the Green's function in Eq.~\ref{eq:Gint},
since here we address nonequilibrium situations.  
The Hartree self-energy is then given by
\begin{equation}
 \Sigma^{\rm H} (t) = \frac{1}{2} U(t)n(t).
\end{equation}
The FLEX prescribes that the electronic self-energy yields
\begin{align}
 \Sigma_{\bolds{k}}^{\rm FLEX} (t,t')=&
 -\frac{i}{N_{k}^{2}} \sum\limits_{\bolds{q}} \Gamma^{ph}_{\bolds{q}}(t,t') G_{\bolds{k}-\bolds{q}}(t,t')  \nonumber \\
 &-\frac{i}{N_{k}^{2}} \sum\limits_{\bolds{q}} \Gamma^{pp}_{\bolds{q}}(t,t')  G_{\bolds{q}-\bolds{k}}(t',t) ,
\end{align}
where the particle-hole (ph) and particle-particle (pp) vertex functions are evaluated by sums of bubble and ladder diagrams as
\begin{align}
 \Gamma^{ph}_{\bolds{q}}(t,t') &= \frac{1}{2} \chi^{c}_{\bolds{q}}(t,t')
 + \frac{3}{2} \chi^{s}_{\bolds{q}}(t,t') -U(t)\chi^{0s}_{\bolds{q}}(t,t')U(t'),\nonumber\\
 \Gamma^{pp}_{\bolds{q}}(t,t') &= \chi^{pp}_{\bolds{q}}(t,t')+U(t)\chi^{0pp}_{\bolds{q}}(t,t')U(t').
\end{align}
In the following, we report our results computed on a lattice with $64 \times 64$ momentum-space discretization.
Due to memory limitation for saving two-time Green's functions, we cease simulating the system after $t_{\rm max}=20$.

\subsection{Observables}Observables which we are interested in analyzing are as follows. 

\begin{figure}[t]
\includegraphics[width=0.47\textwidth]{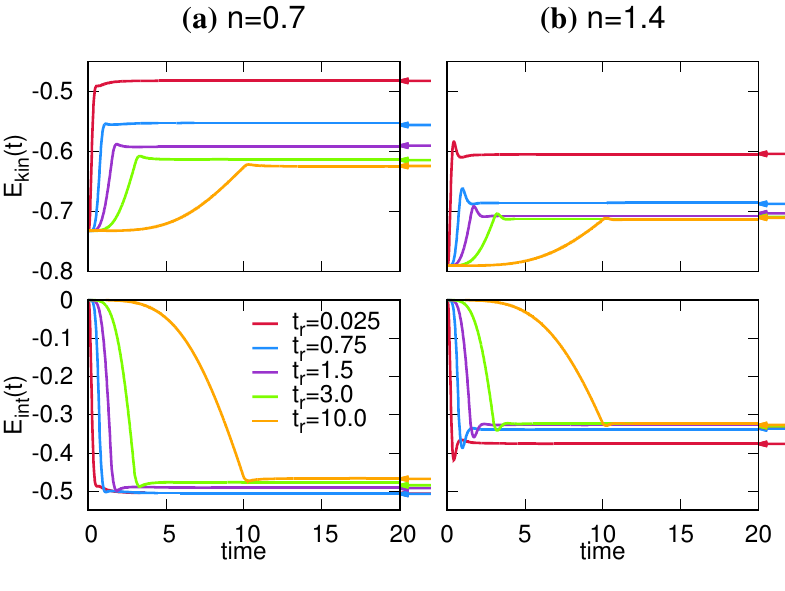}\\[-3mm]
\caption{
Kinetic energy and interaction energy as a function of time for various ramp durations $t_{\rm r}=$ 0.0.25, 0.75, 1.5, 3.0 and 10.0 at fillings $n=$0.7~(a), and 1.4~(b).
Arrows indicate the expected thermal values for each quantity at $t=20$.
}
\label{Fig:Ekin_Eint}
\end{figure}

\paragraph{Kinetic energy} The kinetic energy of the system is calculated as
\begin{equation}
 E_{\rm kin}= \frac{2}{N_{k}^{2}}  \sum\limits_{\bolds{k}} \text{Im} \big[ \varepsilon_{\bolds{k}}G^{<}_{\bolds{k}} \big](t,t).
\end{equation}
\paragraph{Interaction energy} The interaction energy is evaluated as
\begin{equation}
 E_{\rm int}= \frac{2}{N_{k}^{2}}  \sum\limits_{\bolds{k}}\text{Im} \big[ \Sigma^{\rm FLEX}_{\bolds{k}}*G_{\bolds{k}} \big]^{<}(t,t),
\end{equation}
where $*$ again denotes the convolution.  
\paragraph{Total energy} The total energy is 
\begin{equation}
 E_{\rm tot} = E_{\rm kin} + E_{\rm int}.
\end{equation}
The effective temperatures of our relaxed systems is determined from an equilibrium thermal state whose total energy is the same as the driven system at a particular time.

\paragraph{Momentum-dependent distribution function}
The momentum-dependent distribution function is given by
\begin{equation}
 n_{\bolds{k}}= -\text{Im} \big[G_{\bolds{k}} \big]^{<}(t,t).
\end{equation}

\paragraph{Jump of the distribution function}
The jump of the momentum-dependent distribution function is evaluated as
\begin{equation}
 \Delta n_{\bolds{k}_{\rm f}}=n_{\bolds{k}_{\rm f}-\bolds{\delta}} -n_{\bolds{k}_{\rm f}+\bolds{\delta}} ,
\end{equation}
where the momenta of the non-interacting Fermi surface $\bolds{k}_{\rm f}$ satisfies $\varepsilon_{\bolds{k}_{\rm f}}-\mu=0 $. As our system is initially prepared in the noninteracting regime~($U(t=0)=0$), this choice of the Fermi momenta corresponds to the initial Fermi surface. 
Because we are exploring momentum-dependent relaxations, 
we pay particular attention to this quantity at two different Fermi momenta, along the $\Gamma(0,0)- M(\pi,\pi)$, associated with $\bolds{\delta}=  2\pi/N_{k}(1,1)$ and $X(\pi,0)-M(\pi,\pi)$ directions of the Brilloun zone, with $\bolds{\delta} = 2\pi/N_{k}(0,1)$.

\paragraph{Photo-emission spectrum}
The photo-emission spectrum~(PES) of the system probed by a Gaussian pulse~$S(t) =\exp(-t^{2}/2\alpha^{2})$ with the time resolution $\alpha(=3$ here to smear the Fourier artifacts for short-time simulations) 
is given by the 
spectral function,~\cite{Freericks2009}
\begin{eqnarray}
A^{<}(\omega,t)=&\label{eq:PES}\\
-\frac{\mathrm{i}}{2 \pi} \int &{\rm d}t_{1} {\rm d}t_{2} S(t_{1}) S(t_{2})
e^{\mathrm{i} \omega(t_{1}-t_{2})} G^{<}(t+t_{1}, t+t_{2}). \nonumber 
\end{eqnarray} 
To present a time-dependent momentum-resolved PES $A_{\bolds{k}}(\omega,t)$, we substitute the lesser Green's function with the retarded Green's function~($G_{\bolds{k}}^{\rm ret}$).
Calculating the density of state $A_{\rm loc}(\omega,t)$ is then very straightforward as we employ the local retarded Green's function
\begin{equation}
 G_{\rm loc}^{\rm ret}(t,t')= \frac{1}{N_{k}^{2}} \sum\limits_{\bolds{k}} G_{\bolds{k}}^{\rm ret}(t,t'),
\end{equation}
instead of the $G_{\bolds{k}}^{<}$ in Eq.~\ref{eq:PES}.
The obtained spectrum $A$ is broader than the physical spectral function as it is convoluted with the Gaussian function. One should note that, although evaluating the spectral function is straightforward, obtaining a high-resolution spectrum out of our short-time simulations is accompanied by Fourier artifacts. We thus smear these artifacts by convoluting the real-time Green's function with a fairly broad Gaussian filter.

The renormalized dispersion relation~($\widetilde{\varepsilon}_{\bolds{k}}$) at time $t$ is obtained from the position of the quasiparticle peak~($\omega_{\rm qp}$) in $A_{\bolds{k}}(\omega_{\rm qp},t)-U(t)n(t)/2 $ where the factor $U(t)n(t)/2 $ is the Hartree term, see Eq.~\ref{eq:Dysonint}. 

\begin{figure}[t]
\includegraphics{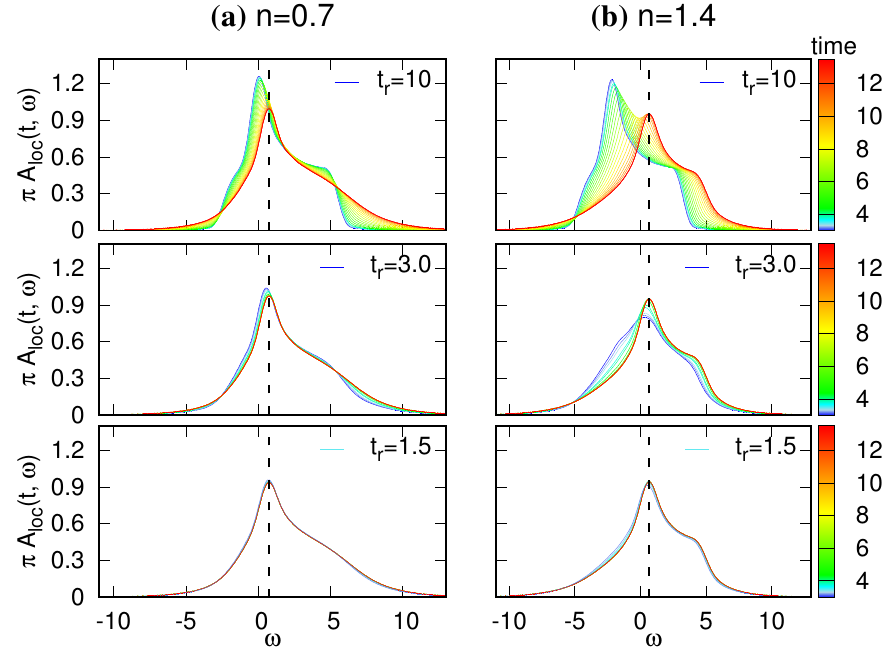}\\[-3mm]
\caption{
The PES for ramp durations $t_{\rm r}=$ 1.5, 3.0, and 10.0 for $U_{\rm f}=3$, $\beta=20$. Vertically aligned panels are at the same fillings, namely $n=$ 0.7~(a), and 1.4~(b). The color coding refers to time. Vertical dashed lines indicate the position of quasiparticle peaks at $\omega=0.75$ for $n=0.7$ and $\omega=0.65$ for $n=1.4$ after the final stage of ramping~($t>t_{\rm r}$). 
}
\label{Fig:Aloc}
\end{figure}
\paragraph{Interacting and noninteracting Fermi surface}
The bare noninteracting Fermi momenta at $U(t=0)=0$ are obtained by solving $\varepsilon_{\bolds{k}}-\mu=0$.
In the presence of electron-electron interaction~($U(t)>0$) momenta on the Fermi surface satisfy $\widetilde{\varepsilon}_{\bolds{k}}-\mu=0$.

\paragraph{Spin, charge, and particle-particle correlation functions} 
The equal-time spin~($s$), charge~($c$), and particle-particle~($pp$) correlation functions are given by
\begin{align}
 \chi^{r}_{\bolds_{k}}(t)=|\text{Im}\left[ {\chi^{r}_{\bolds_{k}}}^{<}(t,t) \right]|,
\end{align}
where ${\chi^{r}_{\bolds_{k}}}^{<}$ denotes the lesser component of the Keldysh susceptibility with the channel index $r \in \{s,c,pp\}$.

\section{Results and discussions}\label{sec:results}

\subsection{Local observables}

Let us first look at the kinetic and potential energies of the hole-doped~($n=0.7$) and electron-doped~($n=1.4$) systems under various ramp durations, $t_{\rm r}$ = 0.025, 0.75, 1.5, 3.0 and 10.0, in Fig.~\ref{Fig:Ekin_Eint}.  
Note here that the nonzero next-neighbor hopping in Eq.~\ref{eq:Hubbard} breaks the particle-hole symmetry, so that, instead of dopings $n$ and $2-n$, we decided to study systems with the chemical potential $\mu_h=-\mu_e$ (which corresponds to $n=0.7$ and $n=1.4$) for the hole- and electron-doped cases. For $n=0.7$ and $n=1.4$ the systems are in the over-doped regimes where the pseudogap phase does not emerge and we are sufficiently far from the Mott insulator.   
Switching on of the Hubbard interaction has a vast effect on the interaction energy naturally for $t<t_{\rm r}$ as seen in lower panels of Fig.~\ref{Fig:Ekin_Eint}~a) and b). The temporal response of the kinetic energy for $t<t_{\rm r}$, similar to the interaction energy, shows a dramatic transient response,
see upper panels of Figs.~\ref{Fig:Ekin_Eint}~(a,b). We can see that the overall change in the kinetic energy strongly depends on whether we sit on the hole-doped or electron-doped side. This is not unexpected for two reasons: (i) our system has the particle-hole asymmetry with different densities of states at the Fermi energy in hole-doped and electron-doped regimes, and (ii) 
an obvious effect in an electron-doped system of smaller numbers of empty sites aside from the thermally generated doublons and holons, which 
forces the electrons to hop to doubly-occupied sites.
In the hole-doped case, hopping to empty sites is allowed, and thus we observe a stronger change in the kinetic energy, 
see upper panel of Fig.~\ref{Fig:Ekin_Eint}(a). Besides, the final values of both kinetic and interaction energy 
agree with the thermal values that are obtained by imposing equal total energy criterion, which indicates local thermalization of these systems.

\begin{figure}[t]
\includegraphics{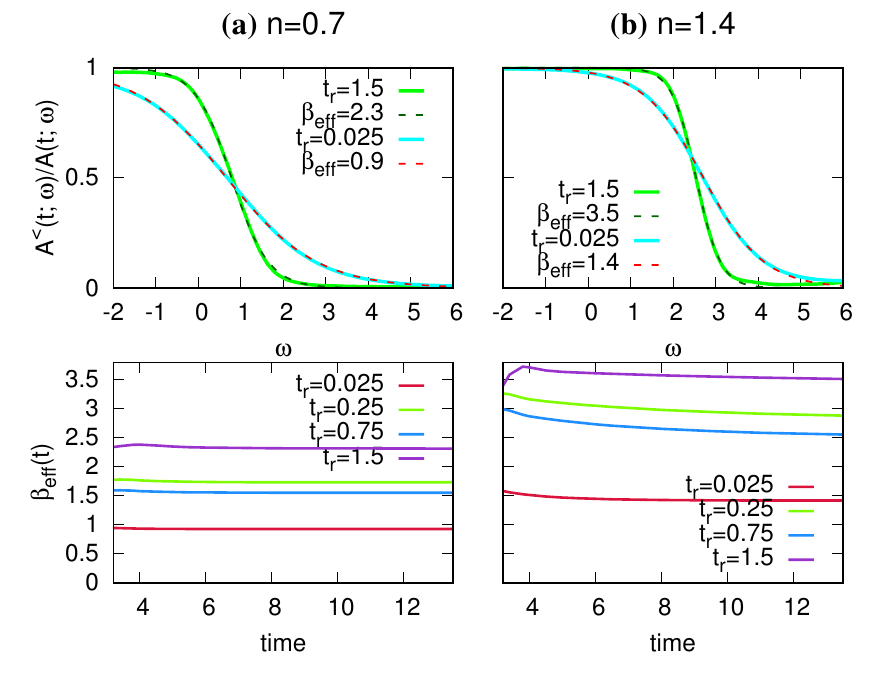}\\[-3mm]
\caption{
For hole-doped~($n=0.7$)~(a) and electron-doped~($n=1.4$)~(b) regimes, the upper panels plot 
the ratio $A^{<}(t,\omega)/A(t,\omega)$ at $t=13.5$ for ramp durations $t_{\rm r}=$  0.025 (yellow solid lines) and 1.5 (cyan solid lines). Dashed lines are fits with the Fermi distribution with the indicated effective inverse temperature, $\beta_{\rm eff}$. Lower panels plot the $\beta_{\rm eff}$ against time for $t_{\rm r}=$ 0.025~(red), 0.25~(green), 0.75~(blue), and 1.5~(purple), again 
for the hole-doped~(left panel) and electron-doped~(right) cases. 
}
\label{Fig:betaeff_Aloc}
\end{figure}

To further corroborate local thermalization of the system, we present the time-evolution of the PES spectrum of systems under various ramp durations, $t_{\rm r}=$ 0.75, 1.5, and 10.0,  in the hole-doped~($n=0.7$) and electron-doped~($n=1.4$) regimes in Fig.~\ref{Fig:Aloc}.  For short ramps, results confirm rapid relaxation of spectrum in all doping regimes, as time-dependent spectrum, plotted with different colors, are hardly distinguishable for $t>6$. For a long ramp~($t_{\rm r}=10$), the evolution of PES is considerable during the preparation period~($t_{\rm r}$) and the evolution quickly terminates shortly after $t_{\rm r}$. It is also notable that at long times when the probe envelope, centered around $t>t_{\rm r}$, overlaps with the initial ramp protocol, the presented evolution of PES is not intrinsic relaxation. At time where local observables became stationary, i.e., $t=13.5$, the spectral density also becomes time-independent, where the position of the quasiparticle peak, independent of the ramp duration, exhibits doping-dependent values as $\omega=0.75$ for $n=0.7$ and $\omega=0.65$ for $n=1.4$, see dashed lines in Fig.~\ref{Fig:Aloc}.

We can further assess the thermal nature of the spectrum by looking at the ratio $A^{<}(t,\omega)/A(t,\omega)$ in the upper panels of Figs.~\ref{Fig:betaeff_Aloc}(a,b), since 
the fluctuation-dissipation theorem dictates that this ratio should be, in a thermal state, equal to the Fermi-Dirac distribution function as
\begin{equation}\label{eq:fluct-dissip}
 \frac{A^{<}(t,\omega)}{A(t,\omega)} =\frac{1}{1+{\rm exp}\left[{\beta_{\rm eff}(t)\big(\omega-\mu_{\rm eff}\big)}\right]}.
\end{equation}
By fitting the spectrum with Eq.~(\ref{eq:fluct-dissip}), we evaluate the inverse of the time-dependent effective temperature~[$\beta_{\rm eff}(t)$]. Consistent with the adiabatic theorem, doped systems with longer ramp durations are effectively colder, see lower panels of Fig.~\ref{Fig:betaeff_Aloc}(a,d). Comparing the effective temperatures of the systems with the same preparation protocol shows that hole-doped systems become hotter than electron-doped systems. We may note that the presented effective temperatures in Fig.~\ref{Fig:betaeff_Aloc} (c,d), derived from a Gaussian-broadened spectrum, are slightly over-estimated, but the trend should be robust.

\subsection{Momentum-dependent observables}

\begin{figure}[t]
\includegraphics[width=0.48\textwidth]{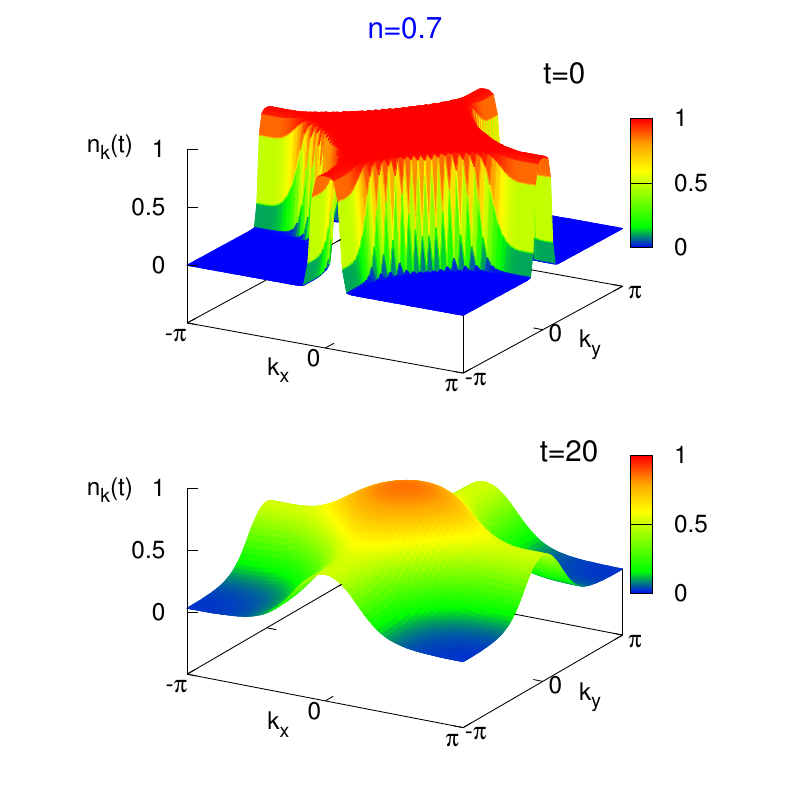}\\[-3mm]
\caption{
For a filling $n=0.7$ the momentum-dependent distribution function at initial~($t=0$; top panel) and final~($t=20$; bottom) times are plotted 
for $U_{\rm f}=3$ and $\beta=30$ and $t_{\rm r}=0.25$.
}
\label{Fig:nk_tq025_holedoped}

\end{figure}
\begin{figure}[t]

\includegraphics{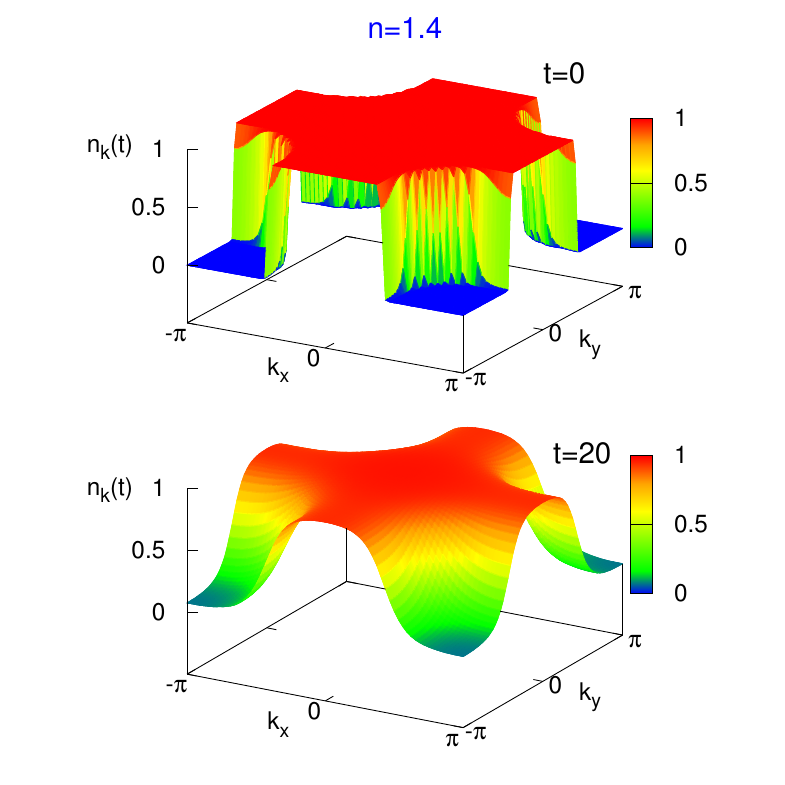}\\[-3mm]
\caption{
For a filling $n=1.4$ the momentum-dependent distribution function at initial~($t=0$; top panel) and final~($t=20$; bottom) times for $U_{\rm f}=3$ and $\beta=30$ and $t_{\rm r}=0.25$.
}
\label{Fig:nk_tq025_eldoped}

\end{figure}

Let us now move on to the momentum-dependent distribution functions in hole-doped~($n=0.7$) and electron-doped~($n=1.4$) systems at initial and final times with $t_{\rm r}=0.25$ in Figs.~\ref{Fig:nk_tq025_holedoped}, and \ref{Fig:nk_tq025_eldoped}, respectively.
A very sharp jump of $n_{k}$ at the Fermi wavenumber at initial times, associated with the Fermi-Dirac distribution at $\beta=20$, is significantly smeared at $t=20$ in both systems. This is consistent with a general intuition that 
occupation of excited states will elevate the effective 
temperature after pumping. 

\begin{figure}[t]
\includegraphics{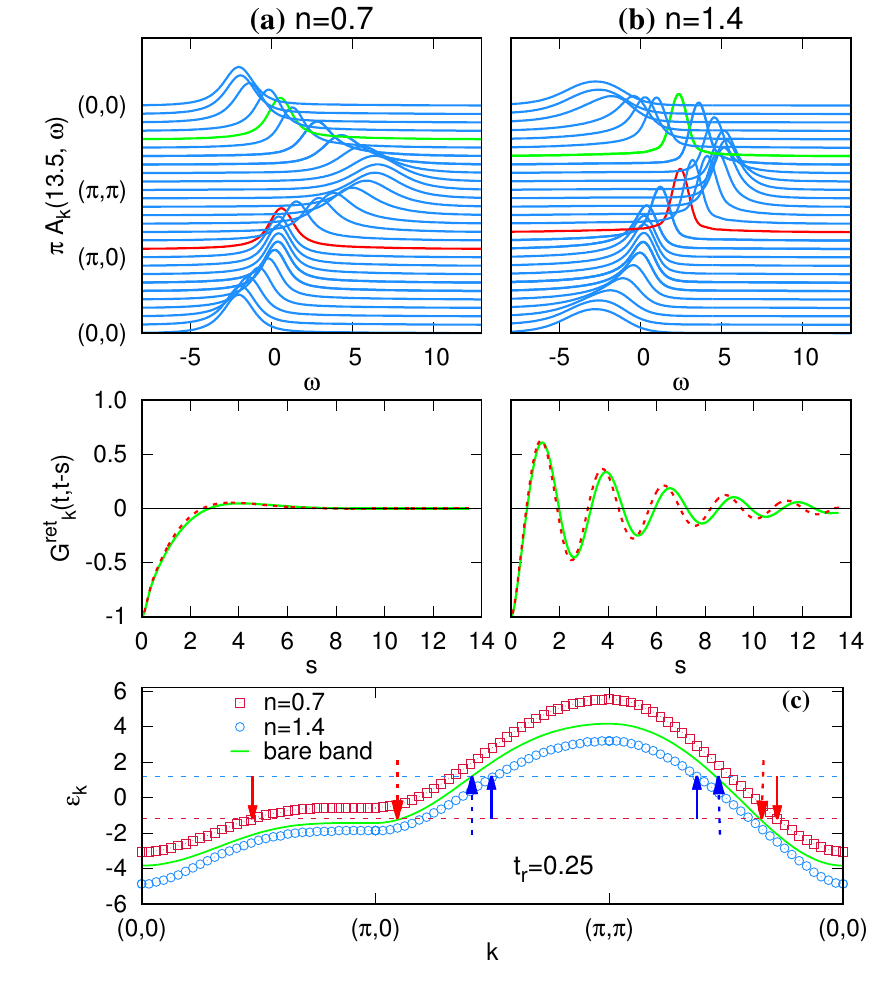}\\[-3mm]
\caption{
Top panels: The PES $A_{\bolds{k}}(t,\omega)$ at $t=13.5$ 
in the hole-doped~($n=0.7$)~(a) and electron-doped~($n=1.4$)~(b) regimes.
The vertical axis is the momentum, with small offsets 
for the spectra for clarity. Red (green) line represents the PES at initial Fermi momenta along $XM$ ($\Gamma M$). 
Middle panels: The retarded Green's functions $G^{\rm ret}_{\bolds{k}}(t,t-s)$ at $t=13.5$ for hole-doped~$(n=0.7$; left panel) and electron-doped~$(n=1.4$; right) cases. Color lines correspond to the spectra with the same colors in upper panels, respectively. 
(c) The dispersion relation for the noninteracting inital band given in Eq.~\eqref{eq:disp}~(solid green line), the renormalized band for the hole-doped~($n=0.7$)~(red open squares) and for the electron-doped~($n=1.4$)~(blue open circles) systems. Dashed~(solid) arrows attached to each dispersion relation indicate the Fermi momenta at $t=0~(t=13.5)$. Horizontal dashed lines represent the chemical potentials for the hole-doped~($n=0.7$)~(red) and the electron-doped~($n=1.4$)~(blue) systems.
}
\label{Fig:Gk}
\end{figure}

We now turn to the upper panels of Fig.~\ref{Fig:Gk}, which plots momentum-resolved PES of both hole-doped~($n=0.7$)~(a) and electron-doped~($n=1.4$)~(b) regimes.  
We can see that the width of the momentum-dependent spectrums close to the Fermi momenta~(red and green lines) is wider for the hole-doped case, which indicates more heating in the hole-doped system than in the electron-doped system. We furthermore observe a doping dependent evolution in $n_{\bolds{k}}$, where the redistribution of occupation in the hole-doped system is so drastic that the occupied region deforms from an open Fermi sea into a closed one, which may be called a ``nonequilibrium Lifshitz transition", see Fig.~\ref{Fig:nk_tq025_holedoped}. In Fig.~\ref{Fig:Gk}~(c) we present the renormalized dispersion relations for the hole-doped~($n=0.7$) and electron-doped~($n=1.4$) systems, extracted from the spectra plotted in upper panels of Fig.~\ref{Fig:Gk}. The band renormalizations are not dramatic as we expect from the weak-coupling physics. The deformation of the Fermi surface in the electron-doped regime~(dashed blue arrows at $t=0$ and solid blue arrows at $t=13.5$) is less considerable than in the hole-doped regime~(dashed red arrows at $t=0$ and solid red arrows at $t=13.5$). Consistent with the presented results for $n_{\bolds{k}}$, see Fig.~\ref{Fig:nk_tq025_holedoped}, the nonequilibrium Lifshitz transition is also evident in these results.

\begin{figure}[t]
\includegraphics{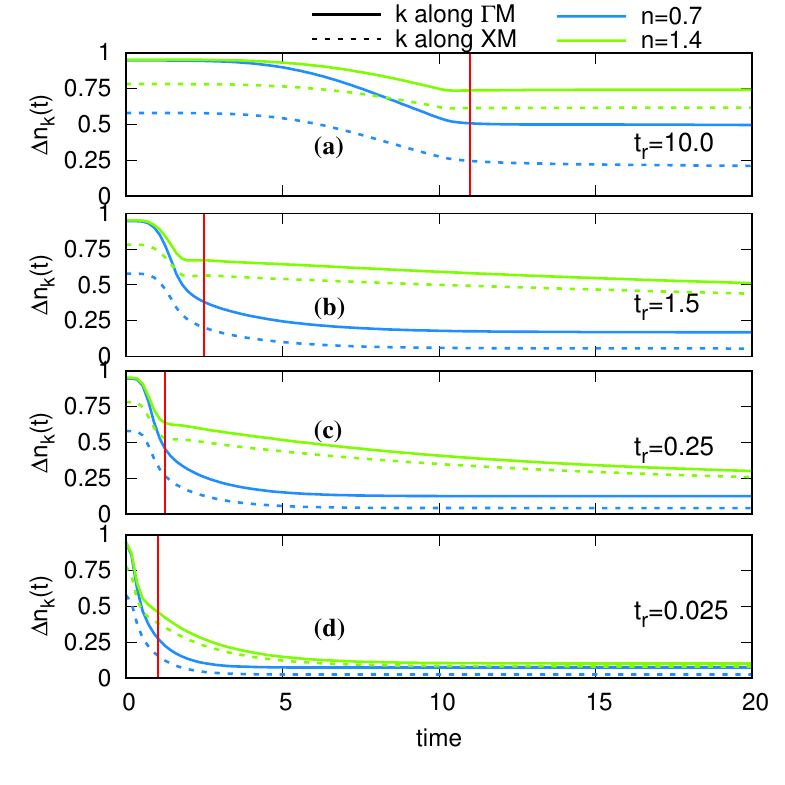}\\[-3mm]
\caption{
The jump, $\Delta n_{\bolds{k}}$, of the momentum-dependent distribution function for ramp durations $t_{\rm r}=$ 10~(a), 1.5~(b), 0.25~(c), and 0.025~(d) at initial Fermi momenta along $\Gamma M$~(solid line) or $XM$~(dashed line) for the hole-doped~($n=0.7$)~(blue) and electron-doped~($n=1.4$)~(red) systems. A vertical red line indicates the transient built-up duration, $t_{\rm  r} + 1/J_1$, 
for each panel.   
}
\label{Fig:deltank}
\end{figure}

To further elaborate the behavior of $n_{\bolds{k}}$ and its doping dependence, Fig.~\ref{Fig:deltank} plots the jump, $\Delta n_{\bolds{k}}$, of the momentum-dependent occupation around the bare noninteracting 
Fermi energy at two Fermi momenta along $\Gamma M$ and $XM$ directions, respectively, in electron-doped, and hole-doped regimes. Note that these Fermi momenta refer to the initial noninteracting Fermi surface for $U(t=0)=0$.

After $t_{\rm r}$, a transient correlation built-up timescale $\sim 1/v_{1}$ is observed, see vertical lines in Fig.~\ref{Fig:deltank}.~\cite{Moeckel2010,Bauer2015} This short-time response is followed by an exponential relaxation dynamics in all the cases. 
The initial preparation protocol and the doping concentrations govern these momentum-dependent evolutions: For the longer ramp durations, the system finds the opportunity to adjust the injected energy and occupies less excited states. 
Namely, the system effectively experiences less heating and consequently the correlation-based relaxation, in contrast to thermal relaxation for small $t_{\rm r}$, rules the dynamics for long ramps. This over-heating picture is indeed consistent with extensively discussed half-filled results for infinite dimensional~\cite{Moeckel2010,Eckstein2009,Eckstein2010,Schiro2010} and two-dimensional~\cite{Tsuji2014,Bauer2015} systems.

\begin{figure}[t]
\includegraphics{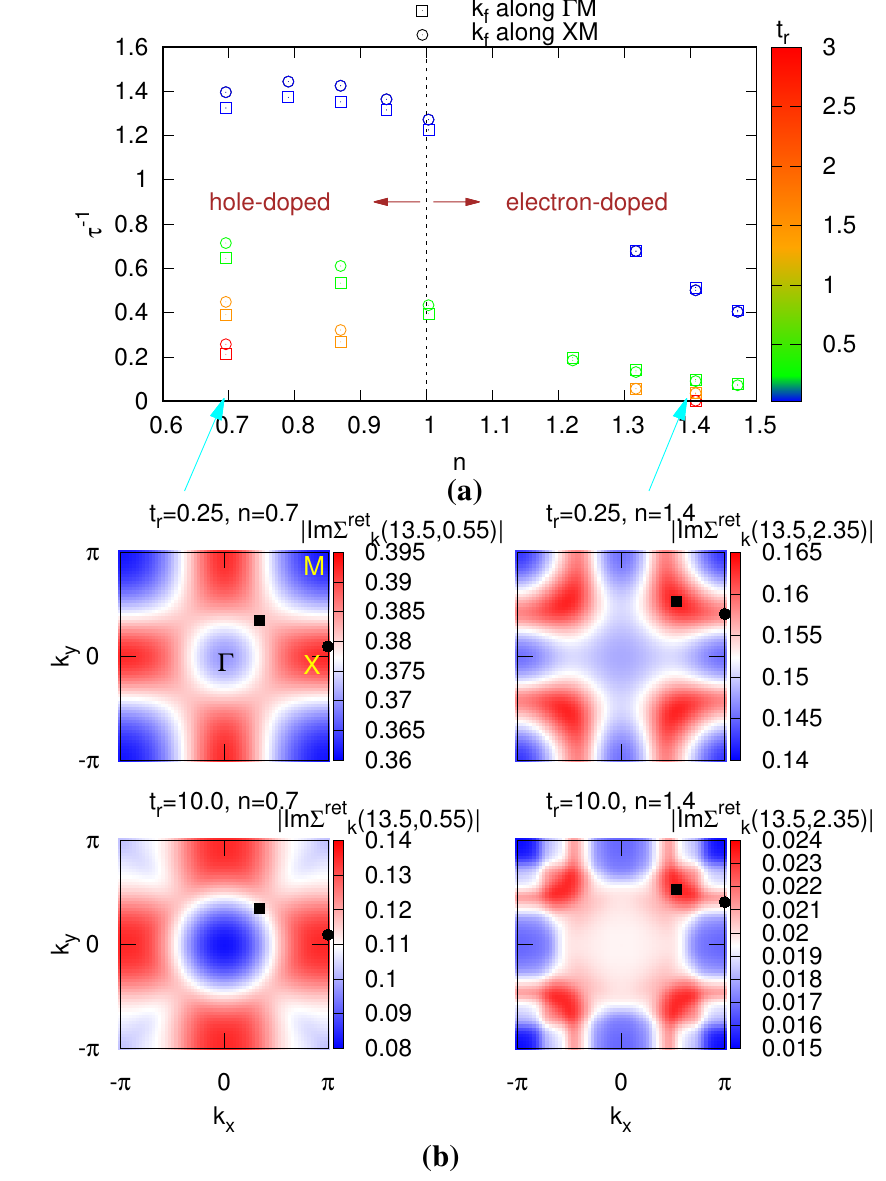}\\[-3mm]
\caption{
(a) The decay rate, $1/\tau$, of $\Delta n_{k}$ at Fermi wave-number at $t=0$ along $\Gamma M$~(squares) and $XM$~(circles) directions in the Brilloun zone, as a function of filling at various ramp durations $t_{\rm r}=$3.0, 1.5, 0.25, and 0.025~(color-coded).  
(b) Color plots of self-energies $|\text{Im}\Sigma^{\rm ret}_{\bolds{k}}(t,\omega_{\rm qp})|$ at $t=13.5$, $\omega_{\rm qp}=0.55$ for the hole-doped~($n=0.7$) system~(left panels), and at $\omega_{\rm qp}=2.35$ for the electron-doped~($n=1.4$) system~(right). The results are for $U_{\rm f}=3$, $\beta=20$, and $t_{\rm r}=0.25$~(middle panles) and $t_{\rm r}=10.0$~(lower panels). Squares~(circles) indicate the Fermi momenta at $t=0$ along $\Gamma M$~($XM$) direction in the Brillouin zone. Note different color codes between different panels.
}
\label{Fig:expdecay_deltank}
\end{figure}

\begin{figure}[t]
\includegraphics{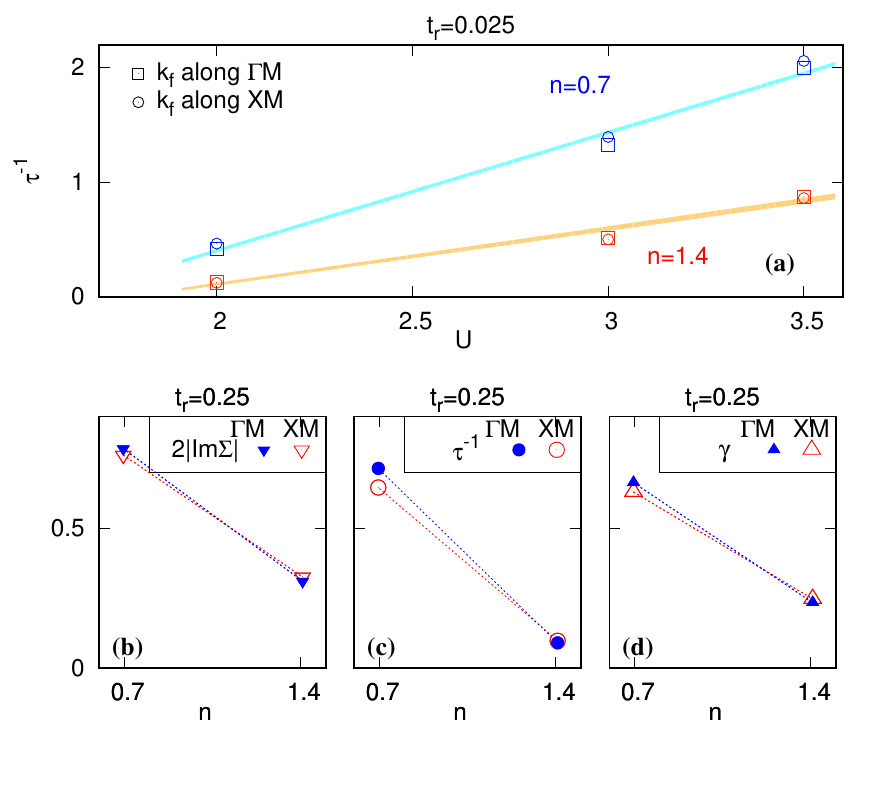}\\[-3mm]
\caption{
(a) The dependence of the relaxation rate on the Hubbard interaction at fillings $n=0.7$~(blue symbols) and $n=1.4$~(red) along $\Gamma M$~(squares) and $XM$~(circles) for $t_{\rm r}=0.025$. Lines are a guide for the eye.
The imaginary part of self-energy~${\rm Im}|\Sigma_{\bolds{k}}|(t,\omega)$~(b), the relaxation rate of $\Delta n_{\bolds{k}}$~(c), and the relaxation rate of the Green's function~(d), at $t=13.5$ and $\omega=\omega_{\rm qp}(\bolds{k})$ as a function of filling for Fermi momenta along $\Gamma M$~(solid symbols) and $XM$~(open) for $t_{\rm r}=0.25$. Dashed lines are for guiding eyes.
}
\label{Fig:relaxation_compare}
\end{figure}

A salient feature is that the relaxation dynamics 
is distinct, in agreement with Fig.~\ref{Fig:deltank}
between the Fermi momenta along the $XM$ and along the $\Gamma M$ 
directions, as previously 
found with nonequilibrium DCA.~\cite{Tsuji2014} 
To quantify this behavior of $\Delta n_{k}$ at finite dopings, we fit our results to an exponential function of the form $ f(t) \propto c_{0}+c_{1}\exp(-t/\tau)  $, where $c_{0}$ and $c_{1}$ are constants, and $\tau$ is the relaxation time. In Fig.~\ref{Fig:expdecay_deltank} we can see that $\tau$ depends on the doping, the Fermi momentum, the Hubbard interaction, and the ramp duration. For long preparation protocols where heating is mitigated, the non-thermal state has a longer lifetime. Furthermore, a stronger electron-electron interaction enhances the scattering rate and thus expedite the relaxation, see Fig.~\ref{Fig:relaxation_compare}(a).  
Also, consistent with the reported undoped DCA results,~\cite{Tsuji2014} the relaxation time is smaller
at the hot-spot, see Fig.~\ref{Fig:expdecay_deltank}(a). 
 The result also shows that the scattering rate 
(inverse relaxation time) decreases with the 
increased density of electrons from hole-doped to electron-doped regimes.
These two features hint distinct structures of the momentum-dependent self-energy in these systems.

We have found that the width of the quasiparticle peaks around the Fermi surface is generically smaller in the electron-doped systems than in the hole-doped ones. The proportionality of this width, in the thermal states in the Fermi liquid theory, to the imaginary part of the retarded self-energy implies smaller incoherent contributions to the scattering rate of the electron-doped systems. This observation then suggests that thermal relaxations are more dominant in the hole-doped systems where the effective local temperature is larger.  
 
To reinforce this conclusion, let us directly look at $\Sigma^{\rm ret}_{\bolds{k}}(t,\omega_{\rm qp})$ both in the hole-doped~($n=0.7$) and electron-doped~($n=1.4$) regimes, where $\omega_{\rm qp}$ denotes the quasiparticle energy, which 
is determined from an almost Lorentzian spectrum, presented in upper panels of Fig.~\ref{Fig:Gk}. In Fig.~\ref{Fig:expdecay_deltank}, lower panels, we plot $|\text{Im}\Sigma^{\rm ret}_{\bolds{k}}(t,\omega_{\rm qp})|$ at $t=13.5$ when the local relaxation for $t_{\rm r}=0.25$ is almost attained, 
with $\omega_{\rm qp}=0.55$ for the hole-doped case~($n=0.7$) or $\omega_{\rm qp}=2.35$ for the electron-doped one~($n=1.4$). The depicted $\omega_{\rm qp}$ for $|\text{Im}\Sigma^{\rm ret}(t,\omega_{\rm qp})|$ in Fig.~\ref{Fig:expdecay_deltank} 
is associated with the quasiparticle energy for $t_{\rm r}=0.25$ 
and at Fermi momenta along the $\Gamma M$ direction in the Brillouin zone, see the green spectra in Fig.~\ref{Fig:Gk}. 
We have also checked that for $t_{\rm r}=10$, the $\omega_{\rm qp}$ is slightly less than for $t_{\rm r}=0.25$.   
 
The anisotropic self-energies for the quasiparticle with the energy $\omega_{\rm qp}$ in hole-doped regime exposes more interacting regions around $(\pi,0)$ in the Brillouin zone.  
In the electron-doped case, we can see the interacting regions shifted to the position of the Fermi surface, see squares~(circles) indicating the Fermi points along $\Gamma M$~($XM$) in Fig.~\ref{Fig:expdecay_deltank}. 
Moreover, in colder systems for larger $t_{\rm r}$, the overall value of the self-energy is smaller, which implies that the quasiparticle has a longer lifetime, in agreement with a slower decay rate of $\Delta n_{k}$ for longer ramps, see Fig.~\ref{Fig:deltank}. This behavior is also evident in the temporal evolution of the two-time retarded Green's function, see middle panels of Fig.~\ref{Fig:Gk}, where the associated electron-doped Green's function exhibits a long-lived oscillation with a slower decay rate.  Let us examine whether the decay rate of quasiparticles characterizing the Green's function describe the inverse relaxation time, $\tau^{-1}$, introduced above. The Fermi liquid theory predicts that for a thermal state the coherent retarded Green's function 
should have the form~\cite{Dasari2018} 
 \begin{equation}
  G_{\bolds{k}}^{\rm ret}(t) \propto -\text{Im}\Big[ e^{-\mathrm{i} \omega_{\rm qp}t} e^{-\gamma_{\bolds{k}}t}\Big],
 \end{equation}
where $\gamma_{\bolds{k}}$ denotes the relaxation rate. We have checked that this asymptotic form matches fairly well with the retarded Green's functions, plotted in middle panels of Fig.~\ref{Fig:Gk}, for $s>t_{\rm r}+1/J_{1}$. In Fig.~\ref{Fig:relaxation_compare}~(d) we display $\gamma_{\bolds{k}_{\rm f}}$~(up-pointing triangles) and $\tau^{-1}_{\bolds{k}_{\rm f}}$~(c) at Fermi momenta~($\bolds{k}_{\rm f}$) along $\Gamma M$~(open symbols) and $XM$~(solid) directions of the Brillouin zone. 

We can see that $\gamma_{\bolds{k}}$ and $\tau^{-1}_{\bolds{k}}$ exhibit similar tendencies against the filling.  
The slight difference between $\gamma_{\bolds k}$ and $\tau^{-1}_{\bolds{k}}$ may be attributed to the remaining non-thermal behavior of the Green's function at $t=13.5$ as well as 
to an error in fitting the short-time data.

 In the weak-coupling limit, the Fermi liquid theory also dictates that for a thermalized system the imaginary part of the retarded self-energy at the quasiparticle energy $\omega_{\rm qp}$ is the inverse quasiparticle lifetime,~\cite{Hodges1971, Allen1987, Sentef2013}
 \begin{equation}
  \tau_{\rm qp}^{-1}(\bolds{k})=- 2 \text{Im} \Sigma^{\rm ret}_{\bolds{k}}(t,\omega_{\rm qp}(\bolds{k})),
 \end{equation}
 where the momentum-dependent quasiparticle energy $\omega_{\rm qp}(\bolds{k})$ at time $t$ indicates the position of the peak in $A_{\bolds{k}}(t,\omega)$, see Fig.~\ref{Fig:Gk}.
To assess whether this relation can also describe the relaxation rate of $\Delta n_{\bolds{k}}$, we compare in Fig.~\ref{Fig:relaxation_compare}~(b) the two quantities, at Fermi momenta along $\Gamma M$ and $XM$ directions, associated with two slightly different $\omega_{\rm qp}$, in hole-doped~($n=0.7$) and electron-doped~($n=1.4$) regimes for $t_{\rm r}=0.25$. Both $\text{Im}\Sigma$ (b) 
and $1/\tau$ (c) decrease as the density of 
electrons are increased. As the self-energy is obtained from the broadened real-time results, a quantitative comparison between the self-energy and the relaxation dynamics is not feasible. Nevertheless, we do notice that $\text{Im}\Sigma$ and $1/\tau$ deviate more from each other in the hole-doped system. This suggests that the Landau quasiparticle picture may not fully describe the relaxation dynamics in the hole-doped system as the relaxation rate at Fermi surface exceeds the associated $\omega_{\rm qp}$; hence the quasiparticle is not well-defined.~\cite{Berthod2013,Sayyad2016}

\begin{figure}
\includegraphics{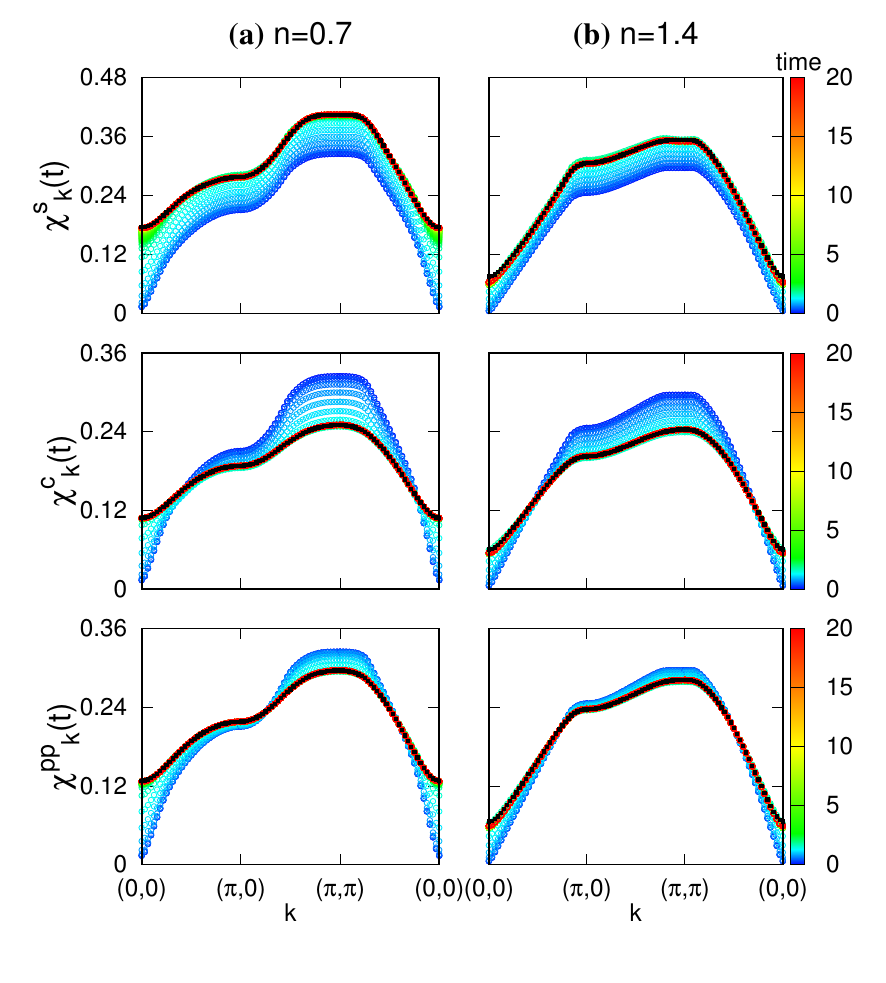}\\[-3mm]
\caption{
Equal-time spin~(upper panels), charge~(middle), particle-particle~(lower) susceptibilities along high-symmetry lines in the Brillouin zone in hole-doped~($n=0.7$)~(a) and electron-doped~($n=1.4$)~(b) regimes for $U_{\rm f}=3$, $\beta=30$ and $t_{\rm r}=1.5$. Black dots represent the associated thermal value.
}
\label{Fig:chis_chic_chipp_tq1}

\end{figure}

Finally to corroborate the thermalization of the doped systems, we present the spin~($\chi_{s}$), charge~($\chi_{c}$), and particle-particle~($\chi_{pp}$) correlation functions in Fig.~\ref{Fig:chis_chic_chipp_tq1}. We find that 
these two-particle quantities exhibit significant 
{\it momentum-dependent evolutions} which saturates toward a thermal value, see black points in Fig.~\ref{Fig:chis_chic_chipp_tq1}. Our results in the hole-doped regime also exhibit different relaxation timescales, in particular, between spins and charges, c.f. upper and middle panels in Fig.~\ref{Fig:chis_chic_chipp_tq1} (a). The effective inverse temperature of the hot electronic state upon thermalization for $t_{\rm r}=1.5$ is $\beta_{\rm eff}=3.53$ in the hole-doped system~($n=0.7$), and $\beta_{\rm eff}=5.13$ in the electron-doped system~($n=1.4$).

\section{Conclusion}~\label{sec:conclusion}

We have studied the nonequilibrium dynamics of doped correlated systems under a global ramp of the repulsive Hubbard interaction. From the thermalization of both electron-doped and hole-doped systems in the weak-coupling regime, we have shown that the effective temperature of carriers driven out of equilibrium is smaller for longer ramps and larger fillings. We have discussed that for long ramps the energy density pumped to the system is smaller, hence the system does not occupy highly excited states. For the doping-dependency of the effective temperature, we have explained that in the electron-doped regime the energy cost for hopping of electrons to different sites is greater, which consequently reduces the probability of forming mobile carriers. More importantly, we have identified momentum-dependent relaxations of the distribution function 
that occurs in a doping-dependent manner.

We have then revealed that, for the hole-doped system, the relaxation rate is distinct between the hot- and cold-spots on the Fermi surface, while the momentum-dependence becomes less pronounced in the electron-doped systems. We have moved on to examine how this would fit with the quasiparticle picture in the Fermi-liquid theory. The present results show that, for the hole-doped system, where thermal relaxation governs the dynamics, quasiparticle is not neccessarily well-defined, whereas in the electron-doped system, which experiences colder environment, is reasonably well described by the Fermi-liquid picture.  
We have finally presented the temporal evolution of equal-time two-body spin, charge and particle-particle correlation functions. We have shown that for small ramps the system is thermalized, which we have corroborated by comparing the final two-body observables with their thermal counterparts. 

Further exploration of the relaxation dynamics of doped systems will require studying the system at lower initial temperatures to allow the ordered phases to appear in thermalization. It is also interesting to investigate this problem in a spin-polarized cases where the spin imbalance will enhance growth of magnetic correlations. Studying multi-band systems with anisotropic
coupling parameters is another intriguing problem, which will shed more light on the formation of spatially modulated orders, which we leave for further studies.

\acknowledgments
The numerical calculations have been carried out at the
ISSP Supercomputer Center at the University of Tokyo.
Sh.S. would like to thank N. Kawashima for providing further computational resources. 
Sh.S. and H.A. acknowledge the support of the ImPACT Program of the Council for Science, Technology and
Innovation, Cabinet Office, Government of Japan (GrantNo. 2015-PM12-05-01) from JST.  H.A. is also supported by JSPS KAKENHI Grant No. JP26247057. M.C. acknowledge support from the H2020 Framework Programme, under ERC Advanced GA No. 692670 ''FIRSTORM``, MIUR PRIN 2015 (Prot. 2015C5SEJJ001) and
SISSA/CNR project ''Superconductivity, Ferroelectricity
and Magnetism in bad metals`` (Prot. 232/2015).

\bibliography{FLEX_Man}

\end{document}